\documentclass[12pt,letterpaper,english,thmsa]{article}
\usepackage[T1]{fontenc}
\usepackage[latin9]{inputenc}
\usepackage{amsmath}
\usepackage{amssymb}
\usepackage{graphicx}
\usepackage{chicago}
\usepackage[nolists,tablesfirst]{endfloat}

\makeatletter

\pdfpageheight\paperheight
\pdfpagewidth\paperwidth

\usepackage{amsfonts}\usepackage{amsthm}

\newtheorem{thm}{Theorem}[section]\newtheorem{cor}{Corollary}[section]\theoremstyle{definition}
\usepackage{babel}

\usepackage{babel}

\makeatother

\usepackage{babel}
\begin{document}

\title{\textbf{Oracle Properties and Finite Sample Inference of the Adaptive Lasso for Time Series Regression Models}}

\author{Francesco Audrino%
\thanks{E-mail: francesco.audrino@unisg.ch.
Address: Bodanstrasse 6, 9000 St.Gallen, Switzerland.%
} \\
University of St.Gallen\and Lorenzo Camponovo%
\thanks{E-mail: lorenzo.camponovo@unisg.ch.
Address: Bodanstrasse 6, 9000 St.Gallen, Switzerland.%
} \\
University of St.Gallen}

\date{December 2013}
\maketitle
\begin{abstract}
We derive new theoretical results on the properties of the adaptive least absolute shrinkage and selection operator (adaptive lasso) for time series regression models. In particular, we investigate the question of how to conduct finite sample inference on the parameters given an adaptive lasso model for some fixed value of the shrinkage parameter. Central in this study is the test of the hypothesis that a given adaptive lasso parameter equals zero, which therefore tests for a false positive. To this end we construct a simple testing procedure and show, theoretically and empirically through extensive Monte Carlo simulations, that the adaptive lasso combines efficient parameter estimation, variable selection, and valid finite sample inference in one step. Moreover, we analytically derive a bias correction factor that is able to significantly improve the empirical coverage of the test on the active variables. Finally, we apply the introduced testing procedure to investigate the relation between the short rate dynamics and the economy, thereby providing a statistical foundation (from a model choice perspective) to the classic Taylor rule monetary policy model.
\mbox{}
\end{abstract}

\section{Introduction}

Recent years have seen a steady increase in the availability of large amounts of economic data. This raises the question of how best to exploit this information to refine benchmark techniques used by the private industry and the academic community. In this study we focus on the question of how to perform consistent variable selection and (finite sample) inference in classical time series regression models with a fixed number of candidate variables and a general error distribution.

To this end, we consider a technique recently introduced in the machine learning community belonging to the group of shrinkage methodologies (that is, following the idea of shrinking to zero the coefficients of the irrelevant variables) that has proven its worth and is becoming increasingly popular: the Least
Absolute Shrinkage and Selection Operator (lasso), introduced by \citeN{T96}, and its refined version known as the adaptive lasso, proposed by \citeN{Z06}.\footnote{We refer to \citeN{BvdG11} for a detailed discussion of the lasso estimator and its generalizations in the cross-sectional setting for independent and identically distributed (iid) variables.}
The main problem with the lasso is that it requires a condition denoted as the {\em irrepresentable condition}, which is essentially a necessary condition for exact recovery of the non-zero coefficients that is much too restrictive in many cases.\footnote{See, for example, \citeN{ZY06} for more details.} Indeed, irrepresentable conditions show that the lasso typically selects too many variables and that so-called false positives are unavoidable. \citeN{Z06} proposed the adaptive lasso to alleviate this problem and to try to reduce the number of false positives. Moreover, the adaptive lasso estimator also fulfills the oracle property in the sense introduced by \citeN{FL01}.

The interest in using lasso-type techniques in general applications as well, such as those in economics and finance, raises the question of how to extend the most advanced theoretical results derived for the iid cross-sectional setting to the time-series setting. Recent papers investigating this question in settings requiring different assumptions and conditions on the number of active variables and on the error distribution include: \citeN{WLT07}, \citeN{HHC08}, \citeN{NR11}, \citeN{SB11}, \citeN{KC12}, \citeN{PS12}, and \citeN{AK12}.

Two studies recently proposed are closely related to our work: \citeN{K12} and \citeN{MM12}.
Each investigates the asymptotic properties of the adaptive lasso estimator in single-equation linear time series models: While \citeN{K12} focuses more on the conditions needed to perform consistent variable selection in stationary and nonstationary autoregressive models using the adaptive lasso with a fixed number of variables, \citeN{MM12} extend the basic linear model investigated in the studies cited above to include exogenous variables, non-Gaussian, conditionally heteroscedastic and possibly time-dependent errors, and a
number of variables (candidate and relevant ones) that is allowed to increase as a function of the sample size. Although some of their results are identical to those of the present study, conditions on the model and proofs are substantially (if not completely) different.

Moreover, in this paper we take the discussion a step further and contribute to the literature on the adaptive lasso along two main dimensions. First, we quantify the bias in finite sample that is incurred when making inference on the active variables in time series regressions by introducing an explicit formula. Second, we show how we can make inference on the inactive variables in the regression. In particular, we introduce a very simple procedure to test the hypothesis that a given parameter is equal to zero, i.e.~that the corresponding variable does not belong to the active set. Our theoretical results show that the adaptive lasso may combine efficient parameter estimation, variable selection, and valid finite sample inference in one step.

To the best of our knowledge, these are new results. They extend the usefulness of the (adaptive) lasso beyond variable selection for performing statistical inference (such as tests of hypotheses and the construction of confidence intervals) in a broad spectrum of applications in all fields dealing with a large amount of iid and time series data.

Some related research on the significance of the active variables in a lasso model has been recently proposed by \citeN{LTTT13}. Nevertheless, their proposed covariance statistic for testing the significance of predictor variables as they enter the active set, along the lasso solution path, differs considerably from the approach we propose in this study. As \citeN{LTTT13} maintain in their discussion section at the end of the paper, the question of how to carry out a significance test for any predictor in the active set given a lasso model at some fixed value of $\lambda_n$ (i.e.~the tuning parameter of the lasso that controls the amount of shrinkage) was left for future research. Finally, a similar approach for the lasso in the Gaussian iid setting has been proposed by \citeN{JM13}.

Using an extensive simulation study based on data generating processes with a different number of variables, error distributions, and number of observations at disposal, we investigate the relevance of the bias correction factor and the effectiveness of the introduced testing procedure. First, results show the importance in finite samples of the bias correction factor for the active variables: the empirical coverages are significantly improved, in particular for variables with small coefficients. Second, although conservative in their construction, tests of the null hypothesis that coefficients of the inactive variables are equal to zero give accurate results, yielding reasonably small proportions of false rejections. Third, if enough data is available, tests on the active variables with small coefficients also produce satisfactory power results.

Finally, we apply the adaptive lasso to a classic problem in financial economics that has been investigated in the academic community for the last twenty years: the relation between interest rates and the state of the economy.\footnote{See, among others, \citeN{CGG00}, \citeN{AP03}, \citeN{DL06}, \citeN{M08}, \citeN{ABW08}, \citeN{R10}, and \citeN{FADG13}.} In particular, we analyze which variables, from a group of macroeconomic and financial indicators, are relevant explaining the short-term interest rate in a simple Taylor rule-type monetary policy model (see \citeANP{T93}, \citeyearNP{T93}, page 202). Not surprisingly, we find that the only predictors belonging to the set of active variables identified by the adaptive lasso are the following three: one-lag past short rate values (which take into account the well-known persistence of the short rate dynamics and act as a proxy for additional macroeconomic, monetary
policy, or even financial variables), an inflation indicator, and the unemployment rate. This result adds a purely statistical foundation to the classic economic intuition driving the Taylor rule, suggesting that the Federal Reserve System (Fed) increases interest rates in times of high inflation, or when employment is above the full employment levels, and decreases interest rates in the opposite situations.

The content of this paper can be summarized as follows: Section 2 introduces the model we are going to consider. Oracle properties of the adaptive lasso for time series regressions are discussed in Section 3. In Section 4, we introduce the statistical testing procedure that can be used to make finite samples inference on both active and inactive variables. Monte Carlo simulation results are shown in Section 5, and the application on the prediction of the short-term interest rate is performed in Section 6. Finally, Section 7 concludes. All proofs of the theorems in the paper are provided in the appendix.

\section{Model and Notation}

Consider the stationary linear regression model
\begin{equation}\label{pr}
Y_t= \sum_{i=1}^{p_1} \rho_{i}^{\ast}Y_{t-i}+\sum_{i=1}^{p_2} \gamma_i^{\ast}W_{i,t}+\sum_{i=1}^{p_3} \beta_i^{\ast}X_{i,t-1}+\epsilon_t,
\end{equation}
where $p_1+p_2+p_3=p<\infty$, $W_{t}=(W_{1,t},\dots,W_{p_2,t})'$ is a vector of covariates at time $t$, $X_{t-1}=(X_{1,t-1},\dots,X_{p_3,t-1})'$ is a vector of regressors at time $t-1$ assumed to predict $Y_t$, $\epsilon_t$ is a zero-mean error term, and $\theta^{\ast}=(\rho_1^{\ast},\dots,\rho_{p_1}^{\ast},\gamma_1^{\ast},\dots,\gamma_{p_2}^{\ast},\beta_1^{\ast},\dots,\beta_{p_3}^{\ast})'$ is the unknown parameter of interest.\footnote{At this point, we intentionally do not introduce restrictions for the correlation structure of the covariates and regressors, except for excluding perfect correlations. Further assumptions on $W_t$ and $X_{t-1}$ are introduced in Theorem \ref{t1} below.} Important examples of linear regression models (\ref{pr}) include: autoregressive models (when $\gamma_i^{\ast}=0$ and $\beta_j^{\ast}=0$, for $i=1,\dots,p_2$ and $j=1,\dots,p_3$); iid
linear regression models (when $\rho_i^{\ast}=0$ and $\beta_j^{\ast}=0$, for $i=1,\dots,p_1$ and $j=1,\dots,p_3$); and predictive regression models (when $\rho_i^{\ast}=0$ and $\gamma_j^{\ast}=0$, for $i=1,\dots,p_1$ and $j=1,\dots,p_2$).

To simplify our notation, we just write $\theta^{\ast}=(\theta_1^{\ast},\dots,\theta_{p}^{\ast})$, i.e., we set $\theta_i^{\ast}=\rho_i^{\ast}$, for $i=1,\dots,p_1$, $\theta_i^{\ast}=\gamma_{i-p_1}^{\ast}$, for $i=p_1+1,\dots,p_1+p_2$, and $\theta_i^{\ast}=\beta_{i-p_1-p_2}^{\ast}$, for $i=p_1+p_2+1,\dots,p_1+p_2+p_3$.
A common way to estimate the unknown parameter $\theta^{\ast}$ relies on the least squares approach. More precisely, we can introduce the least squares estimator $\hat{\theta}_{LS}=(\hat{\theta}_{LS,1},\dots,\hat{\theta}_{LS,p})'$ of $\theta^{\ast}$ defined as
\[
\hat{\theta}_{LS}=arg\min_{\theta}\frac{1}{n}\sum_{t=1}^n (Y_t-\theta'Z_t)^2,
\]
where $Z_t=(Y_{t-1},\dots,Y_{t-p_1},W_t',X_{t-1}')'$ and $n$ denote the sample size. Furthermore, under some regularity conditions and using standard techniques, we can show that
\[
\sqrt{n}(\hat{\theta}_{LS}-\theta^{\ast})\to_{d} N(0,V),
\]
i.e, $\hat{\theta}_{LS}$ is a consistent estimator of $\theta^{\ast}$ with normal limit distribution and covariance matrix $V$.

Let $A=\{i: \theta_i^{\ast}\ne 0\}$ denote the set of the non-zero coefficients, and assume that $\vert A\vert =q<p$. Similarly, Let $\hat{A}_{LS}=\{i: \hat{\theta}_{LS,i}^{\ast}\ne 0\}$. Then, in general $\vert \hat{A}_{LS}\vert =p\ne q$. Thus, in spite of an efficient estimate of the unknown parameter, the least squares approach does not perform variable selection. In the iid context, \citeN{Z06} introduces a lasso procedure which combines both efficient parameter estimation and variable selection in one step. To achieve this objective in time series regression models as well, we extend the lasso method to our setting. More precisely, we introduce the adaptive lasso estimator $\hat{\theta}_{AL}=(\hat{\theta}_{AL,1},\dots,\hat{\theta}_{AL,p})'$ of $\theta^{\ast}$ defined as
\begin{equation}\label{alasso}
\hat{\theta}_{AL}=arg\min_{\theta}\frac{1}{n}\sum_{t=1}^n (Y_t-\theta'Z_t)^2+\frac{\lambda_n}{n}\sum_{i=1}^{p}\lambda_{n,i}\vert \theta_i\vert,
\end{equation}
where $\lambda_n$ is a tuning parameter and $\lambda_{n,i}=1/\vert \hat{\theta}_{LS,i}\vert$.
To simplify the presentation of our results, we consider only the weights $\lambda_{n,i}=1/\vert \hat{\theta}_{LS,i}\vert$ instead of the more general penalizations $\lambda_{n,i}^{(\gamma)}=1/\vert \hat{\theta}_{i}\vert^{\gamma}$ proposed in \citeN{Z06}, where $\gamma>0$ and $\hat{\theta}_{i}$ is a root-$n$ consistent estimator of $\theta_i^{\ast}$, for instance the lasso estimate of $\theta^{\ast}$. However, with some slight modifications we can extend the results in Sections \ref{oracle} and \ref{fin} also to this more general framework.

It is important to note that the penalization of the variables in (\ref{alasso}) also depends on the least squares estimate $\hat{\theta}_{LS}$. In particular, variables with least squares estimates close to zero are more penalized. This property represents a key condition for ensuring valid variable selection, as highlighted in \citeN{Z06} in the iid context.
In the next section, we derive the asymptotic properties of the adaptive lasso for time series regression models.

\section{Oracle Properties of the Adaptive Lasso}\label{oracle}
In the iid context, the adaptive lasso procedure introduced in \citeN{Z06} possesses the so-called oracle properties. More precisely, the adaptive lasso simultaneously performs correct variable selection and provides an efficient estimate of the non-zero coefficients as if only the relevant variables had been included in the model. In the next theorem, we show that the adaptive lasso enjoys these properties in time series regression models as well. Before presenting the main results, first we introduce some notation. Let $\theta^{\ast A}=(\theta^{\ast A}_1,\dots,\theta^{\ast A}_q)'$ denote the sub-vector of the non-zero coefficients of $\theta^{\ast}$. Similarly, let $\hat{\theta}^{A}_{LS}=(\hat{\theta}^{A}_{LS,1},\dots,\hat{\theta}^{A}_{LS,q})'$ and $\hat{\theta}^{A}_{AL}=(\hat{\theta}^{A}_{AL,1},\dots,\hat{\theta}^{A}_{AL,q})'$ denote the least squares and adaptive lasso estimates of $\theta^{\ast A}$. Furthermore, let $V^{A}$ be the asymptotic covariance matrix of $\hat{\theta}^{A}_{LS}$. Finally, let $\hat{A}_{AL}=\{i: \hat{\theta}_{AL,i}\ne 0\}$. The oracle properties of the adaptive lasso are derived in the next theorem.
\begin{thm}\label{t1}
Let $p_1+p_2+p_3=p<\infty$. Assume that $\{Y_t\}$ and $\{Z_t\}$ are stationary processes such that
$\frac{1}{n}\sum_{t=1}^n Z_{t}Z_{t}'\to_{pr}C$, where $C$ is a nonrandom matrix of full rank,
and $\frac{1}{\sqrt{n}}\sum_{t=1}^n \epsilon_{t}Z_{t}\to_d N(0,\Omega)$, for some covariance matrix $\Omega$. If (i) $\lambda_n\to +\infty$ and (ii) $\frac{\lambda_n}{\sqrt{n}}\to 0$, then:
\begin{itemize}
\item[(I)] Variable Selection: $\lim_{n\to\infty}P(\hat{A}_{AL}=A)=1$.
\item[(II)] Limit Distribution:
\[
\sqrt{n}\left(\hat{\theta}_{AL}^{A}-\theta^{\ast A}\right)+\hat{b}_{AL}^{A}\to_d N(0,V^{A}),
\]
where the bias term is given by
\begin{equation*}
\hat{b}_{AL}^{A}=\left(\frac{1}{n}\sum_{i=1}^nZ_t^{A}Z_t'^{A}\right)^{-1}\cdot\left(\frac{\lambda_n}{2\sqrt{n}}\lambda_{n,1}^{A}sign(\hat{\theta}_{AL,1}^{A}),\dots,\frac{\lambda_n}{2\sqrt{n}}\lambda_{n,q}^{A}sign(\hat{\theta}_{AL,q}^{A})\right)',
\end{equation*}
$Z_t^{A}$ is the sub-vector of $Z_t$ for the non-zero coefficients, and $\lambda_{n,i}^{A}=1/\vert \hat{\theta}_{LS,i}^{A}\vert$, $i=1,\dots,q$.
\end{itemize}
\end{thm}

The assumptions in Theorem \ref{t1} are mild conditions that are also required for proving the consistency and deriving the limit distribution of the least squares estimator $\hat{\theta}_{LS}$. Statement (I) of Theorem \ref{t1} establishes that the adaptive lasso performs correct variable selection also in time series settings, i.e. the adaptive lasso asymptotically identifies the sub-vector of the non-zero coefficients of $\theta^{\ast}$. Furthermore, in statement (II) we derive the limit distribution. In particular, we note that the adaptive lasso has the same limit distribution of the least squares estimator. Therefore, the adaptive lasso performs variable selection and efficient parameter estimation in one step.

The oracle properties (I) and (II) discussed above are in line with the results shown in \citeN{K12} and \citeN{MM12}, which are derived using substantially different arguments and proofs.
Moreover, moving beyond those studies, in statement (II) we also provide an explicit formula for the bias term $\hat{b}_{AL}^A$ that is incurred when making inference on the active variables. This term is asymptotically negligible but provides important refinements for finite sample inference, as highlighted in Section \ref{mc} below.

\section{Finite Sample Inference with the Adaptive Lasso}\label{fin}

In the previous section, we derived the asymptotic properties of the adaptive lasso for time series regression models. In particular, we showed that the limit distribution of the estimators of the non-zero coefficients is normal, while those of the zero coefficients collapse to zero. This allows us to use these results to introduce inference on the non-zero coefficients. However, since a priori we do not know the non-zero coefficients $\theta^{\ast A}$, in this context the practical implementation of testing procedures remains unclear.
To deepen our understanding of this issue, suppose that the estimate of the first component of $\theta^{\ast}$ is different from zero, i.e., $\hat{\theta}_{AL,1}\ne0$. Then, we have two cases: (i) $\theta^{\ast}_1\ne0$ or (ii) $\theta^{\ast}_1=0$. If
$\theta^{\ast}_1\ne0$, then by Theorem \ref{t1} the limit distribution of $\hat{\theta}_{AL,1}$ is normal. Thus, we can construct Gaussian confidence intervals for the parameter of interest. On the other hand, if $\theta^{\ast}_1=0$, then by Theorem \ref{t1} we can only conclude that $\hat{\theta}_{AL,1}$ must collapse to zero asymptotically. Since a priori we do not know whether (i) or (ii) is satisfied, it turns out that we also do not know how to make inference on the parameter $\theta^{\ast}_1$.

The aim of this section is to clarify how to introduce valid finite sample inference with the adaptive lasso. In particular, we show that the adaptive lasso may combine efficient parameter estimation, variable selection, and valid finite sample inference in one step.
To achieve this objective, we introduce some notation and terminology in line with \citeN{AG10}. In particular, first we show that the limit distribution of the adaptive lasso is discontinuous in the tuning parameter. Finally, we prove that with an appropriate selection of the critical values, adaptive lasso tests have correct asymptotic size, where the asymptotic size is the limit of the exact size of the test, as defined in (\ref{asy}) below.

To this end,
we slightly change our notation. Let $\hat{\theta}_{AL,\lambda}$ be the adaptive lasso estimate of $\theta^{\ast}$ defined in (\ref{alasso}), where the tuning parameter $0\le \lambda <\infty$ is fixed and does not depend on the sample size $n$. Furthermore, for $i=1,\dots,p$, let $0\le\lambda_{0,i}< \infty$ denote the limit of  $\lambda\vert \hat{\lambda}_{n,i}\vert/\sqrt{n}$, i.e.,
\[
\frac{\lambda\vert \hat{\lambda}_{n,i}\vert}{\sqrt{n}}\to \lambda_{0,i},
\]
as $n\to\infty$. Note that for the non-zero coefficients, $\lambda_{0,i}^{A}=0$.
Then, in the next theorem we derive the limit distribution of $\sqrt{n}(\hat{\theta}_{AL,\lambda}-\theta^{\ast})$.

\begin{thm}\label{t2}
Let $p_1+p_2+p_3=p<\infty$. Assume that $\{Y_t\}$ and $\{Z_t\}$ are stationary processes such that
$\frac{1}{n}\sum_{t=1}^n Z_{t}Z_{t}'\to_{pr}C$, where $C$ is a nonrandom matrix of full rank,
and $\frac{1}{\sqrt{n}}\sum_{t=1}^n \epsilon_{t}Z_{t}\to_d N(0,\Omega)$, for some covariance matrix $\Omega$. Let $0\le \lambda <\infty$. Then,
\[
\sqrt{n}(\hat{\theta}_{AL,\lambda}-\theta^{\ast})\to_d arg\min(R),
\]
where
\begin{equation}
R(u)=-2u'W+u'Cu+\sum_{i=1}^{p}\lambda_{0,i}\vert u_i\vert\label{dist},
\end{equation}
and $W\sim N(0,\Omega)$.
\end{thm}
Note that since for the non-zero coefficients $\lambda_{0,i}^{A}=0$, it turns out that in (\ref{dist}) only the zero coefficients are penalized. Furthermore, when $\lambda=0$, then for $i=1,\dots,p$, $\lambda_{0,i}=0$, and therefore $arg\min(R)=C^{-1}W\sim N(0,V)$, that is the classic least squares case for the full regression. Finally, when $\lambda\to \infty$, the estimates of the zero coefficients collapse to zero.

Suppose that we want to test the null hypothesis $H_{0,i}:\theta^{\ast}_i=\theta^{\ast}_{0i}$ versus the alternative $H_{1,i}:\theta^{\ast}_i\ne\theta^{\ast}_{0i}$, for some $\theta_{0i}^{\ast}\in\mathbb{R}$ and $i\in\{1,\dots,p\}$. To this end, consider the adaptive lasso test statistic $T_{\lambda,i}(\theta_{0i}^{\ast})=\sqrt{n}\vert\hat{\theta}_{AL,\lambda,i}-\theta_{0i}^{\ast}\vert$.
In Theorem \ref{t2}, we establish pointwise convergences that are discontinuous in $\lambda$ and depend on the unknown values $\lambda_{0,i}$, $i=1,\dots,p$. It turns out that because of the lack of uniformity, the limit distribution in (\ref{dist}) can provide very poor approximations of the sampling distribution of the test statistic $T_{\lambda,i}(\theta_{0i}^{\ast})$ under the null hypothesis (see e.g., Andrews and Guggenberger (2010) for more details). To better evaluate the finite sample properties of $T_{\lambda,i}(\theta_{0i}^{\ast})$, it is necessary to study the asymptotic size of the test statistic. Therefore, following \citeN{AG10}, we introduce the exact size and asymptotic size of $T_{\lambda,i}(\theta_{0i}^{\ast})$ as
\begin{eqnarray}
ExSz_n(\theta_{0i}^{\ast})& = & \sup_{\gamma\in\Gamma} P_{H_{0,i},\gamma}(T_{\lambda,i}(\theta_{0i}^{\ast})>z_{1-\alpha}),\nonumber\\
AsySz(\theta_{0i}^{\ast}) &= & \lim \sup_{n\to\infty} ExSz_n(\theta_{0i}^{\ast})\label{asy},
\end{eqnarray}
where the parameter space $\Gamma$ is defined as
$\Gamma = \{ (\theta^{\ast},\lambda,C,\Omega,F) : \theta^{\ast}\in\mathbb{R}^p, 0\le \lambda<\infty, C\in\mathbb{R}^{p\times p}, \Omega\in\mathbb{R}^{p\times p}, det(C)\ne0, det(\Omega)\ne0,$ and $F$ is the joint distribution of the stationary regression model (\ref{pr}) such that
$\frac{1}{n} \sum_{t=1}^n Z_tZ_t'\to_{pr}C$, and $\frac{1}{\sqrt{n}}\sum_{t=1}^n\epsilon_tT_t\to_d N(0,\Omega)\}$, $z_{1-\alpha}$ denote the critical value of the test, and $\alpha\in(0,1)$ is the significance level. As pointed out in \citeN{AG10}, the definition of the asymptotic size incorporates uniformity over $\gamma\in\Gamma$. Therefore, the asymptotic size always ensures a valid approximation of the finite sample size of the test statistic.

\begin{figure}[!h]
\center $ \begin{array}{c}
\includegraphics[height=3in,width=5in]{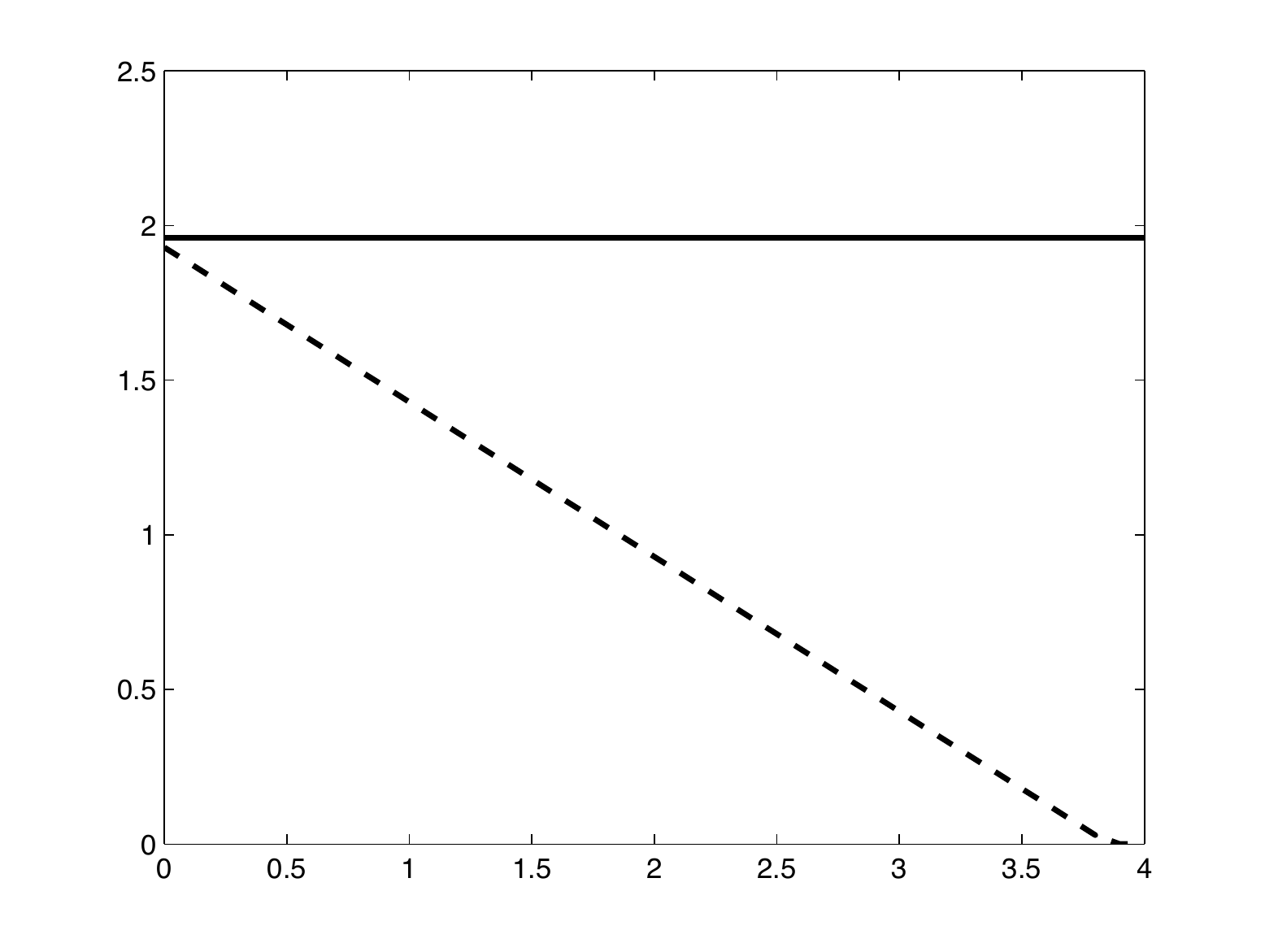}
\end{array}$
\caption{\footnotesize{
 {\bf Quantiles Adaptive Lasso Estimates.}
 We plot the $0.95$-quantiles of the distribution of the random variable $\vert u\vert$, where $u$ minimizes the function $R$ defined in (\ref{dist}).  The horizontal solid line represents the $0.95$-quantile of the distribution of $\vert u\vert$ with $u\sim N(0,1)$. The dashed line represents instead the $0.95$-quantiles of the random variable $\vert u\vert$ for different values of $\lambda_{0,1}\in[0,4]$.}}\label{figure1}
\end{figure}

Using the results in Theorem \ref{t2}, we can show that the adaptive lasso test implies correct asymptotic size. To this end, let $c_{\lambda,i,1-\alpha}$ denote the $1-\alpha$ quantile of the limit distribution of the test statistic $T_{\lambda,i}(\theta_{0i}^{\ast})$. For instance, when $\lambda=0$, then $c_{0,i,1-\alpha}$ is simply the $1-\alpha$ quantile of the random variable $\vert S\vert$, where $S\sim N(0,V_i)$, and $V_i$ is the $i$-th diagonal term of $V$. Using the results in Theorem \ref{t2}, we can easily verify that for all $0\le\lambda<\infty$,
\begin{equation}\label{key}
c_{0,i,1-\alpha}\ge c_{\lambda,i,1-\alpha},
\end{equation}
i.e., the $1-\alpha$ quantile $c_{\lambda,i,1-\alpha}$ is maximized at $\lambda=0$.
To better understand this point, in Figure \ref{figure1} below we plot the $0.95$-quantiles of the distribution of the random variable $\vert u\vert$, where $u$ minimizes the function $R$ defined in (\ref{dist}).

Given the illustrative goal of the figure, we consider the simple case where $u\in\mathbb{R}$ and $C=\Omega=1$. The horizontal solid line in Figure \ref{figure1} represents the $0.95$-quantile of the distribution of $\vert u\vert$ with $u\sim N(0,1)$, i.e, $\lambda_{0,1}=0$ and $c_{0,1,0.95}=1.96$. The dashed line represents instead the $0.95$-quantiles of the random variable $\vert u\vert$ for different values of $\lambda_{0,1}\in[0,4]$ in equation (\ref{dist}). The figure shows that the quantiles are indeed maximized by $c_{0,1,0.95}=1.96$. Then, they decrease almost linearly as $\lambda_{0,1}$ increases. Finally, when $\lambda_{0,1}=4$ the $0.95$-quantile is practically zero.

The result in (\ref{key}) represents the key condition for proving the validity of the finite sample inference with the adaptive lasso. Indeed, consider the adaptive lasso test statistic $T_{\lambda,i}(\theta_{0i}^{\ast})$ with critical value $z_{1-\alpha}=c_{0,i,1-\alpha}$.
Then, using (\ref{key}) it follows trivially that
$AsySz(\theta_{0i}^{\ast}) =\alpha$,
i.e., the adaptive lasso test implies a correct asymptotic size. This result is summarized in the following corollary.
\begin{cor}\label{inf}
Let $p_1+p_2+p_3=p<\infty$. Assume that $\{Y_t\}$ and $\{Z_t\}$ are stationary processes such that
$\frac{1}{n}\sum_{t=1}^n Z_{t}Z_{t}'\to_{pr}C$, where $C$ is a nonrandom matrix of full rank,
and $\frac{1}{\sqrt{n}}\sum_{t=1}^n \epsilon_{t}Z_{t}\to_d N(0,\Omega)$, for some covariance matrix $\Omega$. Let $0\le \lambda <\infty$, and let $\hat{\theta}_{AL,\lambda}$ be the adaptive lasso estimate of $\theta^{\ast}$. Consider the test statistic
$T_{\lambda,i}(\theta_{0i}^{\ast})=\sqrt{n}\vert\hat{\theta}_{AL,\lambda,i}-\theta_{0i}^{\ast}\vert$, with the critical value $c_{0,i,1-\alpha}$.
Then, the asymptotic size of the test of the null hypothesis $H_{0,i}:\theta_i^{\ast}=\theta_{0i}^{\ast}$ versus the alternative $H_{1,i}:\theta_i^{\ast}\ne\theta_{0i}^{\ast}$ satisfies
\begin{equation}
AsySz(\theta_{0i}^{\ast}) =\alpha.\label{cors}
\end{equation}
\end{cor}
Using the result in Corollary \ref{inf} inference based on the adaptive lasso is straightforward.
To test the null hypothesis $H_{0,i}$ at the significance level $\alpha\in(0,1)$ we can simply use the adaptive lasso test statistic $T_{\lambda,i}(\theta_{0i}^{\ast})=\sqrt{n}\vert\hat{\theta}_{AL,\lambda,i}-\theta_{0i}^{\ast}\vert$ with the normal critical value $c_{0,i,1-\alpha}$. As established in (\ref{cors}), this test has correct asymptotic size. This result shows that the adaptive lasso can combine efficient parameter estimation, variable selection, as well as valid finite sample inference in one step.

\section{Monte Carlo}\label{mc}

In this section we use Monte Carlo simulations to study the accuracy of the inference based on the adaptive lasso. In particular, we consider five different settings. To satisfy the assumptions in Theorem 3.1 and as it is customary in the literature, in the Monte Carlo experiments we select the tuning parameter $\lambda_n\in[0,n^{1/4}]$ according to the Bayesian Schwartz Information Criterion (BIC).

\paragraph{Setting 1:} $p_1=p_2=p_3=5$ and $\epsilon_t\sim_{\textrm{iid}} N(0,1)$.

\vspace{0.2cm}\noindent We generate $N=5000$ samples according to model (\ref{pr}) with $p_1=p_2=p_3=5$, $\rho_1^{\ast}=\gamma_1^{\ast}=\beta_1^{\ast}=0.3$,
$\rho_2^{\ast}=\gamma_2^{\ast}=\beta_2^{\ast}=0.1$, and $\rho_i^{\ast}=\gamma_i^{\ast}=\beta_i^{\ast}=0$, for $i=3,4,5$. We consider Gaussian error terms $\epsilon_t\sim_{\textrm{iid}} N(0,1)$. Furthermore, for $i=1,\dots,5$ and $t=1,\dots,n$, let $W_{i,t}\sim_{\textrm{iid}} N(0,1)$ and $X_{i,t-1}\sim_{\textrm{iid}} N(0,1)$. The simulated sample sizes are $n=800$ and $n=1600$.

In a first exercise, we study the accuracy of the inference for the active variables. More precisely, using the results in Theorem \ref{t1}, we construct $0.95$-confidence intervals for the non-zero coefficients $\rho_1^{\ast}=\gamma_1^{\ast}=\beta_1^{\ast}=0.3$, and
$\rho_2^{\ast}=\gamma_2^{\ast}=\beta_2^{\ast}=0.1$. The empirical coverages are summarized in Table \ref{table1}, Panel A. In the first part of Table \ref{table1}, Panel A, we apply the results in Theorem \ref{t1} without the bias term $\hat{b}_{AL}^A$. In contrast, in the bottom part we use instead the bias-corrected limit distribution.
\begin{table}
\begin{center}
{\bf Monte Carlo Simulations: Setting 1}
\end{center}
\begin{center}
Panel A: Empirical coverages for the active variables

$\phantom{P}$

\begin{tabular}{c|cccccc}
\hline
$\phantom{P}$  & $\rho_1^{\ast}=0.3$ & $\rho_2^{\ast}=0.1$ & $\gamma_1^{\ast}=0.3$ & $\gamma_2^{\ast}=0.1$ & $\beta_1^{\ast}=0.3$ & $\beta_2^{\ast}=0.1$  \\
\hline
$n=800$ & $0.9408$ & $0.9234$ & $0.9478$ & $0.9184$ & $0.9404$ & $0.9158$ \\
$n=1600$ & $0.9482$ & $0.9400$ & $0.9476$ & $0.9258$ & $0.9434$ & $0.9284$ \\
\hline
\end{tabular}

$\phantom{P}$

\begin{tabular}{c|cccccc}
\hline
$\phantom{P}$   & $\rho_1^{\ast}=0.3$ & $\rho_2^{\ast}=0.1$ & $\gamma_1^{\ast}=0.3$ & $\gamma_2^{\ast}=0.1$ & $\beta_1^{\ast}=0.3$ & $\beta_2^{\ast}=0.1$  \\
\hline
$n=800$ & $0.9458$ & $0.9496$ & $0.9496$ & $0.9452$ & $0.9424$ & $0.9456$ \\
$n=1600$ & $0.9508$ & $0.9530$ & $0.9500$ & $0.9446$ & $0.9454$ & $0.9472$ \\
\hline
\end{tabular}
\end{center}
\vspace{0.2cm}
\begin{center}
Panel B: Empirical frequencies of rejection of the null hypothesis $H_0$

$\phantom{P}$

\begin{tabular}{c|ccccc}
\hline
$\phantom{P}$  & $\rho_1^{\ast}=0.3$ & $\rho_2^{\ast}=0.1$ & $\rho_3^{\ast}=0$ & $\rho_4^{\ast}=0$ & $\rho_5^{\ast}=0$ \\
\hline
$n=800$ & $1.0000$ & $0.7806$ & $0.0292$ & $0.0302$ & $0.0240$  \\
$n=1600$ & $1.0000$ & $0.9744$ & $0.0248$ & $0.0206$ & $0.0236$ \\
\hline
\end{tabular}

$\phantom{P}$

\begin{tabular}{c|ccccc}
\hline
$\phantom{P}$  & $\gamma_1^{\ast}=0.3$ & $\gamma_2^{\ast}=0.1$ & $\gamma_3^{\ast}=0$ & $\gamma_4^{\ast}=0$ & $\gamma_5^{\ast}=0$ \\
\hline
$n=800$ & $1.0000$ & $0.7318$ & $0.0318$ & $0.0288$ & $0.0268$  \\
$n=1600$ & $1.0000$ & $0.9488$ & $0.0254$ & $0.0280$ & $0.0220$ \\
\hline
\end{tabular}

$\phantom{P}$

\begin{tabular}{c|ccccc}
\hline
$\phantom{P}$  & $\beta_1^{\ast}=0.3$ & $\beta_2^{\ast}=0.1$ & $\beta_3^{\ast}=0$ & $\beta_4^{\ast}=0$ & $\beta_5^{\ast}=0$ \\
\hline
$n=800$ & $1.0000$ & $0.7300$ & $0.0318$ & $0.0288$ & $0.0268$  \\
$n=1600$ & $1.0000$ & $0.9566$ & $0.0240$ & $0.0250$ & $0.0252$ \\
\hline
\end{tabular}
\end{center}
\caption{\footnotesize{The data is generated according to model \eqref{pr} with Gaussian innovations. The simulated sample sizes are $n=800$ and $n=1600$ and the results are based on $N=5000$ simulations.  In Panel A we report the empirical coverages of $0.95$-confidence intervals for the parameters $\rho_1^{\ast}=\gamma_1^{\ast}=\beta_1^{\ast}=0.3$ and
$\rho_2^{\ast}=\gamma_2^{\ast}=\beta_2^{\ast}=0.1$. In the top panel we apply the results in Theorem \ref{t1} without the bias term $\hat{b}_{AL}^A$. In the bottom panel, we use instead the bias-corrected limit distribution. In Panel B we summarize the empirical frequencies of rejection of the null hypothesis $H_{0,i}:\theta_i^{\ast}=0$ versus the alternative $H_{1,i}:\theta_i^{\ast}\ne0$, $i=1,\dots,p$, using the results in Corollary \ref{inf} with significance level $\alpha=0.05$. In the top, second, and bottom panels, we consider
$\rho_i^{\ast}$, $\gamma_i^{\ast}$ and $\beta_i^{\ast}$, $i=1,\dots,5$, respectively.
}}\label{table1}
\end{table}
In the top panel of Table \ref{table1}, Panel A, we note that the adaptive lasso without correction term provides accurate inference for the parameters $\rho_1^{\ast}=\gamma_1^{\ast}=\beta_1^{\ast}=0.3$. Indeed, the empirical coverages are quite close to the nominal coverage probability $0.95$. For instance, when $n=800$ the empirical coverages for $\rho_1^{\ast}$, $\gamma_1^{\ast}$ and $\beta_1^{\ast}$ are $0.9408$, $0.9478$ and $0.9404$, respectively. In contrast, the empirical coverages of the adaptive lasso without correction term for the parameters $\rho_2^{\ast}=\gamma_2^{\ast}=\beta_2^{\ast}=0.1$ are slightly distorted, and tend to be smaller than the nominal coverage probability.
For instance, when $n=800$ the empirical coverages for $\rho_2^{\ast}$, $\gamma_2^{\ast}$ and $\beta_2^{\ast}$ are $0.9234$, $0.9184$ and $0.9158$, respectively.

In the bottom panel of Table \ref{table1}, Panel A, we note that the bias-corrected limit distribution substantially improves the accuracy of the adaptive lasso inference. The empirical coverages using the bias-corrected limit distribution are always closer to the nominal coverage probability than those computed without the bias term $\hat{b}_{AL}^A$. In particular, it is interesting to note that using the bias-corrected distribution the empirical coverages for the parameters
$\rho_2^{\ast}=\gamma_2^{\ast}=\beta_2^{\ast}=0.1$ are very close to $0.95$ as well.
For instance, when $n=800$ the empirical coverages for $\rho_2^{\ast}$, $\gamma_2^{\ast}$ and $\beta_2^{\ast}$ are $0.9496$, $0.9452$ and $0.9456$, respectively. Thus, in this case the difference between empirical coverages and nominal coverage probability is always smaller than $0.005$.
These results show that the bias-corrected limit distribution is particularly important and useful for improving inference on the active variables with small coefficients close to zero.

In a second exercise we study the finite sample power of the introduced adaptive lasso test for the null hypothesis $H_{0,i}:\theta_i^{\ast}=0$ versus the alternative $H_{1,i}:\theta_i^{\ast}\ne0$, $i=1,\dots,p$. Empirical frequencies of rejection of $H_{0,i}$ using the results in Corollary \ref{inf} with significance level $\alpha=0.05$ are reported in Table \ref{table1}, Panel B.\footnote{We test here single hypotheses at significance level $\alpha=0.05$. However, the results in Corollary \ref{inf} can also be used to perform multiple hypothesis testing that allows to control for the familywise error rate (FWER) (see, for example, \citeN{LR05} for more details).}

First, the table shows that for the parameters $\rho_1^{\ast}=\gamma_1^{\ast}=\beta_1^{\ast}=0.3$ we always reject the null hypothesis, for both $n=800$ and $n=1600$. Second, we note that for the parameters $\rho_2^{\ast}=\gamma_2^{\ast}=\beta_2^{\ast}=0.1$, the power of the adaptive lasso test significantly increases as $n$ increases. Thus, in case the coefficients of the active variables are small and close to zero, sufficient data is needed in order for the test to reach high power values. For instance, for $\rho^{\ast}_2=0.1$ the empirical frequencies of rejection of the null hypothesis are $0.7806$ and $0.9744$ for $n=800$ and $n=1600$, respectively.

Finally, the empirical frequencies of rejection for the inactive variables are in a range between $0.02$ and $0.035$. Ideally, the correct value in those cases should equal $\alpha=0.05$, the size of the test. Nevertheless, our results are not surprising and there are two main factors that help explain the values (lower than $\alpha$) we find. First, it is important to remember that the adaptive lasso shrinks some of the coefficients (in particular those of the inactive variables) exactly to zero. In those cases no true asymptotic distribution exists, and therefore the number of non-rejections of the null hypothesis becomes larger. Second, the test we introduced is conservative per construction: this means that we expect to get fewer rejections of the null hypothesis than tests performed under ideal conditions.


 \paragraph{Setting 2:} $p_1=p_2=p_3=5$, $\epsilon_t\sim_{\textrm{iid}} t_5$.

\vspace{0.2cm}\noindent In this setting we study the accuracy of the introduced procedure when dealing with a different error distribution with heavier tails. For this purpose, we generate $N=5000$ samples according to model (\ref{pr}) with the same parameter values and covariates distribution introduced in the first setting.
The only difference is that we assume $\epsilon_t\sim_{\textrm{iid}} t_5$. The simulated sample sizes are $n=800$ and $n=1600$.

As in the previous setting, we analyze both the empirical coverages of $0.95$-confidence intervals for the active variables and the finite sample power of the adaptive lasso test at level $\alpha=0.05$ for the null hypothesis $H_{0,i}:\theta_i^{\ast}=0$ versus the alternative $H_{1,i}:\theta_i^{\ast}\ne0$, $i=1,\dots,p$. 
Results are reported in Table \ref{table6}, Panel A and Panel B, respectively.
\begin{table}
\begin{center}
{\bf Monte Carlo Simulations: Setting 2}
\end{center}
\begin{center}
Panel A: Empirical coverages for the active variables

$\phantom{P}$

\begin{tabular}{c|cccccc}
\hline
  & $\rho_1^{\ast}=0.3$ & $\rho_2^{\ast}=0.1$ & $\gamma_1^{\ast}=0.3$ & $\gamma_2^{\ast}=0.1$ & $\beta_1^{\ast}=0.3$ & $\beta_2^{\ast}=0.1$  \\
\hline
$n=800$ & $0.9442$ & $0.9322$ & $0.9392$ & $0.8984$ & $0.9396$ & $0.8996$ \\
$n=1600$ & $0.9462$ & $0.9418$ & $0.9444$ & $0.9256$ & $0.9422$ & $0.9280$ \\
\hline
\end{tabular}

$\phantom{P}$

\begin{tabular}{c|cccccc}
\hline
   & $\rho_1^{\ast}=0.3$ & $\rho_2^{\ast}=0.1$ & $\gamma_1^{\ast}=0.3$ & $\gamma_2^{\ast}=0.1$ & $\beta_1^{\ast}=0.3$ & $\beta_2^{\ast}=0.1$  \\
\hline
$n=800$ & $0.9464$ & $0.9486$ & $0.9424$ & $0.9196$ & $0.9432$ & $0.9180$ \\
$n=1600$ & $0.9492$ & $0.9516$ & $0.9458$ & $0.9442$ & $0.9482$ & $0.9454$ \\
\hline
\end{tabular}
\end{center}
\vspace{0.2cm}
\begin{center}
Panel B: Empirical frequencies of rejection of the null hypothesis $H_0$

$\phantom{P}$

\begin{tabular}{c|ccccc}
\hline
  & $\rho_1^{\ast}=0.3$ & $\rho_2^{\ast}=0.1$ & $\rho_3^{\ast}=0$ & $\rho_4^{\ast}=0$ & $\rho_5^{\ast}=0$ \\
\hline
$n=800$ & $1.0000$ & $0.7630$ & $0.0384$ & $0.0354$ & $0.0356$  \\
$n=1600$ & $1.0000$ & $0.9674$ & $0.0314$ & $0.0270$ & $0.0292$ \\
\hline
\end{tabular}

$\phantom{P}$

\begin{tabular}{c|ccccc}
\hline
  & $\gamma_1^{\ast}=0.3$ & $\gamma_2^{\ast}=0.1$ & $\gamma_3^{\ast}=0$ & $\gamma_4^{\ast}=0$ & $\gamma_5^{\ast}=0$ \\
\hline
$n=800$ & $1.0000$ & $0.5436$ & $0.0366$ & $0.0388$ & $0.0398$  \\
$n=1600$ & $1.0000$ & $0.8210$ & $0.0346$ & $0.0342$ & $0.0302$ \\
\hline
\end{tabular}

$\phantom{P}$

\begin{tabular}{c|ccccc}
\hline
  & $\beta_1^{\ast}=0.3$ & $\beta_2^{\ast}=0.1$ & $\beta_3^{\ast}=0$ & $\beta_4^{\ast}=0$ & $\beta_5^{\ast}=0$ \\
\hline
$n=800$ & $1.0000$ & $0.5370$ & $0.0394$ & $0.0344$ & $0.0390$  \\
$n=1600$ & $1.0000$ & $0.8248$ & $0.0338$ & $0.0358$ & $0.0334$ \\
\hline
\end{tabular}
\end{center}
\caption{\footnotesize{The data is generated according to model \eqref{pr} with $\epsilon_t\sim_{\textrm{iid}} t_5$. The simulated sample sizes are $n=800$ and $n=1600$ and the results are based on $N=5000$ simulations.  In Panel A we report the empirical coverages of $0.95$-confidence intervals for the parameters $\rho_1^{\ast}=\gamma_1^{\ast}=\beta_1^{\ast}=0.3$ and
$\rho_2^{\ast}=\gamma_2^{\ast}=\beta_2^{\ast}=0.1$. In the top panel we apply the results in Theorem \ref{t1} without the bias term $\hat{b}_{AL}^A$. In the bottom panel, we use instead the bias-corrected limit distribution. In Panel B we summarize the empirical frequencies of rejection of the null hypothesis $H_{0,i}:\theta_i^{\ast}=0$ versus the alternative $H_{1,i}:\theta_i^{\ast}\ne0$, $i=1,\dots,p$, using the results in Corollary \ref{inf} with significance level $\alpha=0.05$.
In the top, second, and bottom panels, we consider $\rho_i^{\ast}$, $\gamma_i^{\ast}$ and $\beta_i^{\ast}$, $i=1,\dots,5$, respectively.
}}\label{table6}
\end{table}
The results in Table \ref{table6}, Panel A, confirm that inference with the adaptive lasso based on the bias-corrected limit distribution substantially outperforms inference based on the limit distribution without finite sample bias correction. For instance, in Table \ref{table6}, Panel A, when $n=1600$ the empirical coverages for $\rho_2^{\ast}$, $\gamma_2^{\ast}$ and $\beta_2^{\ast}$ using the bias-corrected limit distribution are $0.9516$, $0.9442$ and $0.9454$, respectively. In contrast, the empirical coverages for $\rho_2^{\ast}$, $\gamma_2^{\ast}$ and $\beta_2^{\ast}$ without the bias term $\hat{b}_{AL}^A$ are  $0.9418$, $0.9256$ and $0.9280$, respectively. Also in this setting, the adaptive lasso provides a valid statistical tool for testing the null hypothesis $H_{0,i}:\theta_i^{\ast}=0$. Indeed, in Table \ref{table6}, Panel B, we note that the adaptive lasso test always rejects the null hypothesis for $\rho_1^{\ast}=\gamma_1^{\ast}=\beta_1^{\ast}=0.3$, both for $n=800$ and $n=1600$. Furthermore, the power of the test for small coefficients close to zero can be moderate, in particular when not enough data is available. However, the power of the test substantially increases as $n$ increases. In particular, in Table \ref{table6}, Panel B, we observe that when $n=1600$ the empirical frequencies of rejection for $\rho_2^{\ast}$, $\gamma_2^{\ast}$ and $\beta_2^{\ast}$ are larger than $0.8$. Finally, the proportion of false rejections for the inactive variables is again in most cases around $0.035$, close to the level $\alpha$ of the test.

 \paragraph{Setting 3:} $p_1=p_2=p_3=5$, and GARCH error terms.

\vspace{0.2cm}\noindent In this setting we study the accuracy of the adaptive lasso procedure with heteroscedastic error terms. For this purpose, we generate $N=5000$ samples according to model (\ref{pr}) with the same parameter values and covariate distributions introduced in the first two settings.
For the error terms we assume the following GARCH representation
\begin{eqnarray}
\epsilon_t & = & \sqrt{h_t}e_t,\label{gar1}\\
h_{t} & = & 0.1+0.7h_{t-1}+0.1h_{t-1}e_{t-1}^2,\label{gar2}
\end{eqnarray}
where $e_t\sim_{\textrm{iid}} t_5$.
The simulated sample sizes are $n=800$ and $n=1600$.

We perform the same type of analysis as in the previous two settings. Results are reported in Table \ref{table8}, Panel A and Panel B, respectively.
\begin{table}
\begin{center}
{\bf Monte Carlo Simulations: Setting 3}
\end{center}
\begin{center}
Panel A: Empirical coverages for the active variables

$\phantom{P}$

\begin{tabular}{c|cccccc}
\hline
  & $\rho_1^{\ast}=0.3$ & $\rho_2^{\ast}=0.1$ & $\gamma_1^{\ast}=0.3$ & $\gamma_2^{\ast}=0.1$ & $\beta_1^{\ast}=0.3$ & $\beta_2^{\ast}=0.1$  \\
\hline
$n=800$ & $0.9346$ & $0.9168$ & $0.9436$ & $0.9060$ & $0.9418$ & $0.9088$ \\
$n=1600$ & $0.9430$ & $0.9306$ & $0.9438$ & $0.9270$ & $0.9454$ & $0.9274$ \\
\hline
\end{tabular}

$\phantom{P}$

\begin{tabular}{c|cccccc}
\hline
   & $\rho_1^{\ast}=0.3$ & $\rho_2^{\ast}=0.1$ & $\gamma_1^{\ast}=0.3$ & $\gamma_2^{\ast}=0.1$ & $\beta_1^{\ast}=0.3$ & $\beta_2^{\ast}=0.1$  \\
\hline
$n=800$ & $0.9386$ & $0.9288$ & $0.9480$ & $0.9308$ & $0.9434$ & $0.9316$ \\
$n=1600$ & $0.9450$ & $0.9442$ & $0.9452$ & $0.9456$ & $0.9492$ & $0.9472$ \\
\hline
\end{tabular}
\end{center}
\vspace{0.2cm}
\begin{center}
Panel B: Empirical frequencies of rejection of the null hypothesis $H_0$

$\phantom{P}$

\begin{tabular}{c|ccccc}
\hline
  & $\rho_1^{\ast}=0.3$ & $\rho_2^{\ast}=0.1$ & $\rho_3^{\ast}=0$ & $\rho_4^{\ast}=0$ & $\rho_5^{\ast}=0$ \\
\hline
$n=800$ & $1.0000$ & $0.5460$ & $0.0438$ & $0.0380$ & $0.0420$  \\
$n=1600$ & $1.0000$ & $0.7872$ & $0.0406$ & $0.0338$ & $0.0360$ \\
\hline
\end{tabular}

$\phantom{P}$

\begin{tabular}{c|ccccc}
\hline
  & $\gamma_1^{\ast}=0.3$ & $\gamma_2^{\ast}=0.1$ & $\gamma_3^{\ast}=0$ & $\gamma_4^{\ast}=0$ & $\gamma_5^{\ast}=0$ \\
\hline
$n=800$ & $1.0000$ & $0.6640$ & $0.0310$ & $0.0338$ & $0.0328$  \\
$n=1600$ & $1.0000$ & $0.9096$ & $0.0292$ & $0.0262$ & $0.0268$ \\
\hline
\end{tabular}

$\phantom{P}$

\begin{tabular}{c|ccccc}
\hline
  & $\beta_1^{\ast}=0.3$ & $\beta_2^{\ast}=0.1$ & $\beta_3^{\ast}=0$ & $\beta_4^{\ast}=0$ & $\beta_5^{\ast}=0$ \\
\hline
$n=800$ & $1.0000$ & $0.6686$ & $0.0372$ & $0.0316$ & $0.0356$  \\
$n=1600$ & $1.0000$ & $0.9092$ & $0.0268$ & $0.0324$ & $0.0288$ \\
\hline
\end{tabular}
\end{center}
\caption{\footnotesize{The data is generated according to model \eqref{pr} with the error term following a GARCH process with a $t_5$ distributed innovation. The simulated sample sizes are $n=800$ and $n=1600$ and the results are based on $N=5000$ simulations.  In Panel A we report the empirical coverages of $0.95$-confidence intervals for the parameters $\rho_1^{\ast}=\gamma_1^{\ast}=\beta_1^{\ast}=0.3$ and
$\rho_2^{\ast}=\gamma_2^{\ast}=\beta_2^{\ast}=0.1$. In the top panel we apply the results in Theorem \ref{t1} without the bias term $\hat{b}_{AL}^A$. In the bottom panel, we use instead the bias-corrected limit distribution. In Panel B we summarize the empirical frequencies of rejection of the null hypothesis $H_{0,i}:\theta_i^{\ast}=0$ versus the alternative $H_{1,i}:\theta_i^{\ast}\ne0$, $i=1,\dots,p$, using the results in Corollary \ref{inf} with significance level $\alpha=0.05$.
In the top, second and bottom panels, we consider $\rho_i^{\ast}$, $\gamma_i^{\ast}$ and $\beta_i^{\ast}$, $i=1,\dots,5$, respectively.
}}\label{table8}
\end{table}
The results in Table \ref{table8} clearly confirm the same findings arise in the previous two settings.
First, the adaptive lasso with the bias-corrected term provides empirical coverages very close to the nominal coverage probability. Second, the adaptive lasso provides a valid statistical tool for testing the null hypothesis $H_{0,i}:\theta_i^{\ast}=0$ also in presence of heteroscedastic error terms.


\paragraph{Setting 4:} $p_1=1$, $p_2=20$, $p_3=0$ and $\epsilon_t\sim_{\textrm{iid}} N(0,1)$.

\vspace{0.2cm}\noindent In this setting, we investigate the performance of the adaptive lasso procedure in the case of a persistent time series regression with contemporaneous covariates of heterogeneous predictive strengths.  We generate $N=5000$ samples according to model (\ref{pr}) with $p_1=1$, $p_2=20$, $p_3=0$, $\rho_1^{\ast}=0.9$, $\gamma_1^{\ast}=0.6$,
$\gamma_2^{\ast}=0.5$, $\gamma_3^{\ast}=0.4$,
$\gamma_4^{\ast}=0.3$, $\gamma_5^{\ast}=0.2$,
$\gamma_6^{\ast}=0.1$, and $\gamma_i^{\ast}=0$, for $i=7,\dots,20$. We consider Gaussian error terms $\epsilon_t\sim_{\textrm{iid}} N(0,1)$. Furthermore, for $i=1,\dots,20$ and $t=1,\dots,n$, let $W_{i,t}\sim_{\textrm{iid}} N(0,1)$. The simulated sample sizes are $n=800$ and $n=1600$.

As in the previous settings, we perform the same two exercises. We focus first on the empirical coverages and we construct $0.95$-confidence intervals for the parameters $\rho_1^{\ast}=0.9$, $\gamma_1^{\ast}=0.6$,
$\gamma_2^{\ast}=0.5$, $\gamma_3^{\ast}=0.4$,
$\gamma_4^{\ast}=0.3$, $\gamma_5^{\ast}=0.2$,
$\gamma_6^{\ast}=0.1$. In the second exercise we investigate the finite sample power of the adaptive lasso test for the null hypothesis $H_{0,i}:\theta_i^{\ast}=0$ versus the alternative $H_{1,i}:\theta_i^{\ast}\ne0$, $i=1,\dots,p$. The significance level is $\alpha=0.05$. Table \ref{table4}  summarizes the results.
\begin{table}
\begin{center}
{\bf Monte Carlo Simulations: Setting 4}
\end{center}
\begin{center}
Panel A: Empirical coverages for the active variables

$\phantom{P}$

\begin{tabular}{c|ccccccc}
\hline
$\phantom{P}$  & $\rho_1^{\ast}=0.9$  & $\gamma_1^{\ast}=0.6$ & $\gamma_2^{\ast}=0.5$ & $\gamma_3^{\ast}=0.4$ & $\gamma_4^{\ast}=0.3$ & $\gamma_5^{\ast}=0.2$ & $\gamma_6^{\ast}=0.1$  \\
\hline
$n=800$ & $0.9442$ & $0.9392$ & $0.9456$ & $0.9422$ & $0.9462$ & $0.9352$ & $0.9094$  \\
$n=1600$ & $0.9454$ & $0.9464$ & $0.9412$ & $0.9470$ & $0.9428$ & $0.9400$ & $0.9248$  \\
\hline
\end{tabular}

$\phantom{P}$

\begin{tabular}{c|ccccccc}
\hline
$\phantom{P}$  & $\rho_1^{\ast}=0.9$  & $\gamma_1^{\ast}=0.6$ & $\gamma_2^{\ast}=0.5$ & $\gamma_3^{\ast}=0.4$ & $\gamma_4^{\ast}=0.3$ & $\gamma_5^{\ast}=0.2$ & $\gamma_6^{\ast}=0.1$  \\
\hline
$n=800$ & $0.9438$ & $0.9410$ & $0.9492$ & $0.9434$ & $0.9532$ & $0.9442$ & $0.9450$  \\
$n=1600$ & $0.9460$ & $0.9474$ & $0.9428$ & $0.9474$ & $0.9450$ & $0.9434$ & $0.9454$  \\
\hline
\end{tabular}
\end{center}
\vspace{0.2cm}
\begin{center}
Panel B: Empirical frequencies of rejection of the null hypothesis $H_0$

$\phantom{P}$

\begin{tabular}{c|ccccccc}
\hline
$\phantom{P}$  & $\rho_1^{\ast}=0.9$  & $\gamma_1^{\ast}=0.6$ & $\gamma_2^{\ast}=0.5$ & $\gamma_3^{\ast}=0.4$ & $\gamma_4^{\ast}=0.3$ & $\gamma_5^{\ast}=0.2$ & $\gamma_6^{\ast}=0.1$  \\
\hline
$n=800$ & $1.0000$ & $1.0000$ & $1.0000$ & $1.0000$ & $1.0000$ & $1.0000$ & $0.7278$  \\
$n=1600$  & $1.0000$ & $1.0000$ & $1.0000$ & $1.0000$ & $1.0000$ & $1.0000$ & $0.9572$  \\
\hline
$\phantom{P}$  & $\gamma_7^{\ast}=0$ & $\gamma_8^{\ast}=0$ & $\gamma_9^{\ast}=0$ & $\gamma_{10}^{\ast}=0$ & $\gamma_{11}^{\ast}=0$ & $\gamma_{12}^{\ast}=0$ & $\gamma_{13}^{\ast}=0$  \\
\hline
$n=800$ & $0.0332$ & $0.0300$ & $0.0284$ & $0.0334$ & $0.0270$ & $0.0320$ & $0.0322$  \\
$n=1600$  & $0.0240$ & $0.0200$ & $0.0258$ & $0.0204$ & $0.0212$ & $0.0202$ & $0.0222$  \\
\hline
$\phantom{P}$  & $\gamma_{14}^{\ast}=0$ & $\gamma_{15}^{\ast}=0$ & $\gamma_{16}^{\ast}=0$ & $\gamma_{17}^{\ast}=0$ & $\gamma_{18}^{\ast}=0$ & $\gamma_{19}^{\ast}=0$ & $\gamma_{20}^{\ast}=0$  \\
\hline
$n=800$ & $0.0308$ & $0.0300$ & $0.0278$ & $0.0336$ & $0.0350$ & $0.0264$ & $0.0316$  \\
$n=1600$  & $0.0224$ & $0.0254$ & $0.0228$ & $0.0210$ & $0.0228$ & $0.0210$ & $0.0218$  \\
\hline
\end{tabular}
\end{center}
\caption{\footnotesize{The data is generated according to model \eqref{pr} with Gaussian innovations. The simulated sample sizes are $n=800$ and $n=1600$ and the results are based on $N=5000$ simulations.  In Panel A we report the empirical coverages of $0.95$-confidence intervals for the active variables. In the top panel we apply the results in Theorem \ref{t1} without the bias term $\hat{b}_{AL}^A$. In the bottom panel, we use instead the bias-corrected limit distribution. In Panel B we summarize the empirical frequencies of rejection of the null hypothesis $H_{0,i}:\theta_i^{\ast}=0$ versus the alternative $H_{1,i}:\theta_i^{\ast}\ne0$, $i=1,\dots,p$, using the results in Corollary \ref{inf} with significance level $\alpha=0.05$.
}}\label{table4}
\end{table}
Table \ref{table4}, Panel A, reports the empirical coverages for the active variables. As in the previous settings, in this case as well we can observe that the adaptive lasso provides valid inference for the parameters of interest. In particular, the empirical coverages of the adaptive lasso with the bias-corrected term are always very close to the nominal coverage probability. Indeed, the difference between empirical coverages and nominal coverage probability is smaller than $0.01$ both for $n=800$ and $n=1600$. On the other hand, when the parameter of interest is close to zero, the empirical coverages of the adaptive lasso without bias-corrected term can be slightly distorted. For instance, when $n=800$ the empirical coverage for $\gamma_6^{\ast}=0.1$ is $0.9094$.

Furthermore, also in this setting we show in Table \ref{table4}, Panel B, that the adaptive lasso provides a valid statistical tool for testing the null hypothesis $H_{0,i}:\theta_i^{\ast}=0$ versus the alternative $H_{1,i}:\theta_i^{\ast}\ne0$, $i=1,\dots,p$. In particular, the adaptive lasso test always rejects the null hypothesis for $\rho_1^{\ast}$ and $\gamma_i^{\ast}$, $i=1,\dots,5$,
when $n=800$ and $n=1600$. Moreover, when $n=1600$ the power of the adaptive lasso test for $\gamma_6^{\ast}$ is larger than $0.95$. Finally, the proportion of false rejections for the inactive variables is in the range between $0.02$ and $0.035$, close to the level of the test $\alpha$.

\paragraph{Setting 5:} $p_1=1$, $p_2=20$, $p_3=0$, GARCH errors terms and correlated regressors.

\vspace{0.2cm}\noindent In this final setting, we study the accuracy of the adaptive lasso procedure with persistent time series regressions and correlated covariates with heterogenous predictive strengths.
To this end, we generate $N=5000$ samples according to model (\ref{pr}) with the same parameter values as in the previous setting.
We assume for $i=1,\dots,20$ and $t=1,\dots,n$, $W_{i,t}\sim N(0,1)$, with pairwise contemporaneous correlations selected from the set $\{-0.5;0;0.5;0.9\}$. 
Finally, for the error terms
we assume the GARCH representation defined in (\ref{gar1})-(\ref{gar2}).
The simulated sample sizes are $n=800$ and $n=1600$.
This setting reproduces closely the characteristics of the empirical data we analyze in Section \ref{emp} below.

Results for the same analysis as the one performed in the previous setting are summarized in Table \ref{table5}. Once again, results and conclusions from the simulations are similar to those already discussed for the other settings.
\begin{table}
\begin{center}
{\bf Monte Carlo Simulations: Setting 5}
\end{center}
\begin{center}
Panel A: Empirical coverages for the active variables

$\phantom{P}$

\begin{tabular}{c|ccccccc}
\hline
$\phantom{P}$  & $\rho_1^{\ast}=0.9$  & $\gamma_1^{\ast}=0.6$ & $\gamma_2^{\ast}=0.5$ & $\gamma_3^{\ast}=0.4$ & $\gamma_4^{\ast}=0.3$ & $\gamma_5^{\ast}=0.2$ & $\gamma_6^{\ast}=0.1$  \\
\hline
$n=800$ & $0.9456$ & $0.9358$ & $0.9354$ & $0.9396$ & $0.9380$ & $0.9340$ & $0.9022$  \\
$n=1600$ & $0.9506$ & $0.9438$ & $0.9444$ & $0.9478$ & $0.9432$ & $0.9420$ & $0.9180$  \\
\hline
\end{tabular}

$\phantom{P}$

\begin{tabular}{c|ccccccc}
\hline
$\phantom{P}$  & $\rho_1^{\ast}=0.9$  & $\gamma_1^{\ast}=0.6$ & $\gamma_2^{\ast}=0.5$ & $\gamma_3^{\ast}=0.4$ & $\gamma_4^{\ast}=0.3$ & $\gamma_5^{\ast}=0.2$ & $\gamma_6^{\ast}=0.1$  \\
\hline
$n=800$ & $0.9460$ & $0.9434$ & $0.9428$ & $0.9434$ & $0.9396$ & $0.9420$ & $0.9158$  \\
$n=1600$ & $0.9504$ & $0.9472$ & $0.9498$ & $0.9516$ & $0.9486$ & $0.9520$ & $0.9474$  \\
\hline
\end{tabular}
\end{center}
\vspace{0.2cm}
\begin{center}
Panel B: Empirical frequencies of rejection of the null hypothesis $H_0$

$\phantom{P}$

\begin{tabular}{c|ccccccc}
\hline
$\phantom{P}$  & $\rho_1^{\ast}=0.9$  & $\gamma_1^{\ast}=0.6$ & $\gamma_2^{\ast}=0.5$ & $\gamma_3^{\ast}=0.4$ & $\gamma_4^{\ast}=0.3$ & $\gamma_5^{\ast}=0.2$ & $\gamma_6^{\ast}=0.1$  \\
\hline
$n=800$ & $1.0000$ & $1.0000$ & $1.0000$ & $1.0000$ & $1.0000$ & $0.9808$ & $0.5286$  \\
$n=1600$  & $1.0000$ & $1.0000$ & $1.0000$ & $1.0000$ & $1.0000$ & $1.0000$ & $0.8130$  \\
\hline
$\phantom{P}$  & $\gamma_7^{\ast}=0$ & $\gamma_8^{\ast}=0$ & $\gamma_9^{\ast}=0$ & $\gamma_{10}^{\ast}=0$ & $\gamma_{11}^{\ast}=0$ & $\gamma_{12}^{\ast}=0$ & $\gamma_{13}^{\ast}=0$  \\
\hline
$n=800$ & $0.0290$ & $0.0348$ & $0.0338$ & $0.0284$ & $0.0344$ & $0.0284$ & $0.0328$  \\
$n=1600$  & $0.0242$ & $0.0252$ & $0.0276$ & $0.0226$ & $0.0242$ & $0.0262$ & $0.0276$  \\
\hline
$\phantom{P}$  & $\gamma_{14}^{\ast}=0$ & $\gamma_{15}^{\ast}=0$ & $\gamma_{16}^{\ast}=0$ & $\gamma_{17}^{\ast}=0$ & $\gamma_{18}^{\ast}=0$ & $\gamma_{19}^{\ast}=0$ & $\gamma_{20}^{\ast}=0$  \\
\hline
$n=800$ & $0.0322$ & $0.0348$ & $0.0290$ & $0.0292$ & $0.2882$ & $0.0306$ & $0.0286$  \\
$n=1600$  & $0.0238$ & $0.0258$ & $0.0300$ & $0.0282$ & $0.0256$ & $0.0262$ & $0.0284$  \\
\hline
\end{tabular}
\end{center}
\caption{\footnotesize{The data is generated according to model \eqref{pr} with GARCH error terms and correlated covariates. The simulated sample sizes are $n=800$ and $n=1600$ and the results are based on $N=5000$ simulations.  In Panel A we report the empirical coverages of $0.95$-confidence intervals for the active variables. In the top panel we apply the results in Theorem \ref{t1} without the bias term $\hat{b}_{AL}^A$. In the bottom panel, we use instead the bias-corrected limit distribution. In Panel B we summarize the empirical frequencies of rejection of the null hypothesis $H_{0,i}:\theta_i^{\ast}=0$ versus the alternative $H_{1,i}:\theta_i^{\ast}\ne0$, $i=1,\dots,p$, using the results in Corollary \ref{inf} with significance level $\alpha=0.05$.
}}\label{table5}
\end{table}

In sum, the Monte Carlo results in different settings confirm that the adaptive lasso provides a valid approach for testing the null hypothesis $H_{0,i}:\theta_i^{\ast}=0$, $i=1,\dots,p$.




\section{Empirical Illustration\label{emp}}
We consider the relation between the short-term interest rate and the state of the economy in a Taylor-type monetary policy model, i.e.~a linear regression model for the short rate that has as possible regressor candidates all macroeconomic and financial variables such as inflation, unemployment, industrial production, or monetary variables. The macroeconomic data was downloaded from the Federal Reserve Bank of Philadelphia and is part of the database called {\em Real-Time Data Set for Macroeconomists}, which consists of vintages of the most relevant macroeconomic variables. In our study we use the vintage available at the end of 2013. The time period under consideration goes from January 1959 to December 2012, for a total of 648 monthly observations. We collected the data for 15 macroeconomic variables,\footnote{Some of the original variables were excluded from the analysis because of almost perfect collinearity with other variables we consider.} including:
\begin{itemize}
\item[-] {\em Price level indices}: Produced Price Index (PPPI);
\item[-] {\em Monetary and financial}: M1 Money Stock (M1), M2 Money Stock (M2), Monetary Base (BASEBASA), Total Reserves (TRBASA), Nonborrowed Reserves (NBRBASA);
\item[-] {\em Industrial production and capacity utilization}: Capacity Utilization Rate Manufacturing (CUM), Industrial Production Index Total (IPT);
\item[-] {\em Housing}: Housing Starts (HSTARTS);
\item[-] {\em Labor market}: Nonfarm Payroll Employment (EMPLOY), Aggregate Weekly Hours Goods-Producing (HG), Civilian Labor Force (LFC), Participation Rate, Constructed (LFPART), Civilian Noninstitutional Population (POP), Unemployment Rate (RUC).
\end{itemize}
The dependent variable is the US 3-month short rate. Data was downloaded from different sources: from 1959 to the end of 1969 from MacCulloch, from 1970 to the end of 1981 from Fama-Bliss (CRSP), and the final period until the end of 2013 from the Board of Governors of the Federal Reserve System (3-Month Treasury constant maturity rate). Where needed, all variables were seasonally adjusted. To take into account the time series dynamics of the short rate, we included the first lagged short rate value as predictor in the regression. Thus, we have $p_1=1, p_2=15, p_3=0$ in model (\ref{pr}).

In Table \ref{table3}, we report the adaptive lasso point estimates. The tuning parameter $\lambda_n$ is selected according to BIC.
\begin{table}
\begin{center}
\begin{tabular}{lccc}
\hline
Variables & AL Estimate & Standard Errors & LS Estimate \\
\hline
US $3$-month one-lag & $\hphantom{-}0.942921^{***}$ & $0.031752$ & $\hphantom{-}0.920980^{***}$ \\
\hline
Produced Price Index & $\hphantom{-}0.014345^{*\hphantom{**}}$ & $0.007907$ & $\hphantom{-}0.016774^{**\hphantom{*}}$ \\
\hline
M1 Money Stock & $0$ & $0.000328$ & $-0.000630^{*\hphantom{**}}$ \\
M2 Money Stock & $-0.000097\hphantom{^{***}}$ & $0.000180$ & $-0.000684^{***}$ \\
Monetary Base &  $\hphantom{-}0.000202\hphantom{^{***}}$ & $0.001547$ & $\hphantom{-}0.003487^{**\hphantom{*}}$ \\
Total Reserves & $0$ & $0.001796$ & $-0.002998^{*\hphantom{**}}$ \\
Nonborrowed Reserves & $0$ & $0.000348$ & $\hphantom{-}0.000131\hphantom{^{***}}$ \\
\hline
Capacity Utilization Rate Manufacturing & $0$ & $0.015339$ & $-0.024990\hphantom{^{***}}$ \\
Industrial Production Index Total & $0$ & $0.023639$ & $-0.014557\hphantom{^{***}}$ \\
\hline
Housing Starts & $\hphantom{-}0.000122\hphantom{^{***}}$ & $0.000089$ & $\hphantom{-}0.000184^{**\hphantom{*}}$ \\
\hline
Nonfarm Payroll Employment & $-0.000033\hphantom{^{***}}$ & $0.000044$ & $-0.000132^{***}$ \\
Aggregate Weekly Hours Goods- Producing & $\hphantom{-}0.016553\hphantom{^{***}}$ & $0.015850$ & $\hphantom{-}0.036516^{**\hphantom{*}}$ \\
Civilian Labor Force & $\hphantom{-}0.000032\hphantom{^{***}}$ & $0.000038$ & $\hphantom{-}0.000109^{***}$ \\
Participation Rate, Constructed & $-0.034452\hphantom{^{***}}$ & $0.027265$ & $-0.066091^{**\hphantom{*}}$ \\
Civilian Noninstitutional Population & $0$ & $0.000023$ & $\hphantom{-}0.000030\hphantom{^{***}}$ \\
Unemployment Rate & $-0.169115^{*\hphantom{**}}$ & $0.092235$ & $-0.252584^{***}$ \\
\hline
\end{tabular}
\end{center}
\caption{\footnotesize{{\bf Taylor rule monetary policy model for the short rate: Adaptive lasso point estimation and inference.} We report the adaptive lasso estimates (column 2), the standard errors (column 3) and the full least squares estimates (column 4) for the empirical analysis introduced in Section \ref{emp}. The dependent variable is the US 3-month short rate. The period under investigation ranges from January 1959 to December 2012, for a total of 648 monthly observations. Asterisks $^{*}$ ,$^{**}$ ,$^{***}$ denote significance at the 10$\%$, 5$\%$ and 1$\%$ level, respectively.}}\label{table3}
\end{table}
Applying Corollary \ref{inf} we test the null hypothesis $H_{0,i}:\theta_{i}^{\ast}=0$, $i=1,\dots,16$, ($15$ regressors and the first lagged short rate as predictors).
As shown in Table \ref{table3}, column 2, the only variables that are significantly different from zero using the adaptive lasso testing procedure are the lagged short rate, the Producer Price Index, and the Unemployment Rate. It is interesting to see that there are other variables with adaptive lasso estimates different from zero. Without the use of the testing procedure introduced in this study we would not have been able to classify them as false positives.


This result is not surprising. Indeed, the predictors we find to be statistically significant and to belong to the active set of variables identified by the adaptive lasso procedure are those also commonly thought to be economically relevant in the Taylor rule monetary policy model for the short rate. According to this rule, the central bank sets the nominal short-term interest rate, $r_t$, based on the following equation

\begin{equation*}\label{eq:TaylorRule1}
r_{t} = \gamma_{0} +  \rho r_{t-1} + \gamma_{\pi}\, \pi_{t}  + \gamma_g \, g_t  + \varepsilon^{r}_{t},
\end{equation*}
where $\pi_t$ denote inflation, $g_t$ is the output gap, and $\varepsilon^{r}_t$ is a sequence of independent and normally distributed innovations with mean zero and variance $\sigma_{r}^2$. Thus, our result adds a purely statistical foundation (from the viewpoint of variable choice in the regression) to this economically intuitive rule. Moreover, the sign and (partially) the magnitude of the coefficients of the active variables are in line with the literature, that is a positive relation between inflation and the short rate, a negative relation between unemployment and the short rate, and a high persistence of the short rate dynamics.\footnote{See, for example, \citeN{FADG13} and the references therein.}

It is important to highlight that this result would not have been possible without the theory for testing null hypotheses of the type $H_{0,i}:\theta_i^{\ast}=\theta_{0i}^{\ast}$ versus the alternative $H_{1,i}:\theta_i^{\ast}\ne\theta_{0i}^{\ast}$, for some $\theta_{0i}^{\ast}\in\mathbb{R}$, developed in this study. In fact we would have found more active variables using both the adaptive lasso and the classical full least squares estimates (whose results are summarized in column 4 of Table \ref{table3}), completely losing the economic intuition behind the Taylor rule and rendering the interpretation of the results very difficult.

\section{Conclusions \label{sec:conclusion}}

We presented new theoretical and empirical results on the finite sample and asymptotic properties of the adaptive lasso in time series regression models. We extended previous results presented in the literature along two main lines: (i) computing analytically a bias correction term for doing finite sample inference on the active variables in the adaptive lasso, and (ii) introducing a simple, conservative, but effective testing procedure for the null hypothesis that a parameter is equal to zero in the adaptive lasso model with a fixed amount of shrinkage.

Through extensive Monte Carlo simulations with a changing number of candidate variables, different error distributions, different sample sizes, and different correlation structures among the covariates, we showed the accuracy of the testing procedure in finite sample. Moreover, testing our procedure in a more involved simulation experiment where we relaxed the assumption of iid errors, we also empirically confirmed the theoretical results and showed that the methodology is robust against this kind of deviation from the standard setting. This result is not surprising and confirms the recent findings and discussions in \citeN{MM12} and \citeN{K12}.

Finally, we investigated the implications of the new testing procedure in an empirical application concerning the relation between the short-term interest rate dynamics and the (macro)economy. To this end, we considered a Taylor rule monetary policy model, where we let the adaptive lasso choose from a number of macroeconomic and financial predictors the relevant ones to put in the active set. We then tested using the new procedure to see whether all remaining active variables had a corresponding coefficient significantly different from zero. In contrast with the full least squares approach on all variables, the only variables with a coefficient different than zero identified by our testing procedure were an inflation indicator, the unemployment rate, and the one-lagged past short rate. We interpreted this result as a statistical confirmation of the Taylor rule.

Our theoretical results are general and can be applied to a broad spectrum of iid and time series applications, in particular when the researcher has to do variable selection and inference among many candidate variables. Classic examples are realized volatility modeling, excess returns or inflation prediction. Moreover, in light of the theoretical results proved in this study, an alternative way of conducting finite sample inference can be envisaged. In the spirit of the recent works proposed by \citeN{CL11} and \citeN{CL13}, we plan to develop bootstrap simulation techniques that can be applied to the (adaptive) lasso to perform finite sample testing of the resulting parameters. Finally, we plan to investigate whether the theory we introduced can be generalized to the case where the number of variables is increasing with the sample size and/or applications dealing with more variables than observations. This is left for future research.

\newpage

\section*{Appendix: Proofs of the Theorems}

\noindent {\bf Proof of Theorem \ref{t1}:} First we derive the limit distribution of the adaptive lasso. In particular, we adopt the same argument as in the proof of Theorem 2 in \citeN{Z06}. Finally, we use this result to prove (I) Variable Selection and to compute the bias term in (II) Limit Distribution.

Let
\[
R_n(u)= \sum_{t=1}^n \left[ (\epsilon_t-u'Z_t/\sqrt{n})^2-\epsilon_t^2\right]+\lambda_n\sum_{i=1}^{p} \lambda_{n,i}\left[ \vert \theta_i^{\ast}+u_i/\sqrt{n}\vert -\vert \theta_i^{\ast}\vert\right].
\]
Note that $R_n$ is minimized at $\sqrt{n}(\hat{\theta}_{AL}-\theta^{\ast})$. Furthermore, we know that
\[
\sum_{t=1}^n \left[ (\epsilon_t-u'Z_t/\sqrt{n})^2-\epsilon_t^2\right] \to_d -2u'W+u'Cu,
\]
where $W\sim N(0,\Omega)$. Now, consider the limit of the second term $\lambda_n\sum_{i=1}^{p} \lambda_{n,i}\left[ \vert \theta_i^{\ast}+u_i/\sqrt{n}\vert -\vert \theta_i^{\ast}\vert\right] $. If $\theta_i^{\ast}\ne 0$, then $\lambda_{n,i}\to \vert \theta_i^{\ast}\vert^{-1}$, and consequently $\lambda_n \lambda_{n,i}\left[ \vert \theta_i^{\ast}+u_i/\sqrt{n}\vert -\vert \theta_i^{\ast}\vert\right]\to 0$. If $\theta_i^{\ast} = 0$, then $
\vert \theta_i^{\ast}+u_i/\sqrt{n}\vert -\vert \theta_i^{\ast}\vert=u_i/\sqrt{n}$, and furthermore $\lambda_n\lambda_{n,i}=\lambda_nC$, where $C/\sqrt{n}=O_p(1)$. Let $R(u)$ denote the limit of $R_n(u)$. Then, we can conclude that
\[
R(u) = \left\{ \begin{array}{ll}
-2u_A'W^{A}+u_A'C^{A}u_A & \textrm{if } u_i=0, \textrm{ for } i\notin A,\\
\infty & \textrm{otherwise,}
\end{array} \right.
\]
where $W^{A}\sim N(0,\Omega^{A})$ and $\Omega^{A}$ is the sub-matrix of $\Omega$ for the non-zero coefficients. Note that $R_n$ is convex, and the unique minimum of $R$ is $((C^{A})^{-1}W^{A},0)'$. Therefore, by \citeN{G94} it follows that
\begin{eqnarray*}
\sqrt{n}\left(\hat{\theta}_{AL}^{A}-\theta^{\ast A}\right) & \to_d & N(0,V^{A}),\\
\sqrt{n}\hat{\theta}_{AL}^{A^c}& \to_d & 0,
\end{eqnarray*}
where $\hat{\theta}_{AL}^{A^c}$ denote the adaptive lasso estimate of the zero coefficients $\theta^{\ast A^c}$ of $\theta^{\ast}$.

Using this result, we can prove (I) Variable Selection.
We adopt the same argument as in the proof of Lemma 5 in \citeN{FP04}.
Let
\[
Q(\theta)=\frac{1}{n}\sum_{t=1}^n (Y_t-\theta'Z_t)^2+\frac{\lambda_n}{n}\sum_{i=1}^{p}\lambda_{n,i}\vert \theta_i\vert.
\]
With some abuse of notation, we write $\theta^{\ast}=(\theta'^{\ast A},\theta'^{\ast A^c})'$. We show that with probability tending to $1$, for any $\hat{\theta}^{A}$ satisfying $\Vert \hat{\theta}^{A}-\theta^{\ast A}\Vert =O_p(1/\sqrt{n})$ and any constant $C$,
\[
Q((\hat{\theta}'^A,0')')=min_{\Vert \hat{\theta}^{A^c}\Vert\le C/\sqrt{n}}Q((\hat{\theta}'^A,\hat{\theta}'^{A^c})').
\]
To this end, for $j\notin A$ consider
\begin{eqnarray*}
\frac{\partial Q(\theta)}{\partial\theta_{j}}  & = &  -\frac{2}{n}\sum_{i=1}^n(Y_t-\theta'Z_t)Z_{t}^{(j)}+\frac{\lambda_n}{n}\lambda_{n,j}sign(\theta_{j})\\
& = & J_1+J_2,
\end{eqnarray*}
where $Z_t^{(j)}$ denote the $j$-component of the vector $Z_t$.
Note that $J_1=O_p(1/\sqrt{n})$, while the dominant term is $J_2$, since  (i) $\lambda_n\to +\infty$ and (ii) $\frac{\lambda_n}{\sqrt{n}}\to 0$. Thus, the sign of $\theta_{j}$ determines the sign of
$\frac{\partial Q(\theta)}{\partial\theta_{j}}$. More precisely, we have
\begin{eqnarray*}
 & \frac{\partial Q(\theta)}{\partial\theta_{j}} < 0, & \textrm{when $-\epsilon_n<\theta_{j}<0$}.\\
 & \frac{\partial Q(\theta)}{\partial\theta_{j}} > 0, & \textrm{when $0<\theta_{j}<\epsilon_n$}.
\end{eqnarray*}
This concludes the proof of (I) Variable Selection.

Finally, using these results, we can focus on (II) Limit Distribution and also derive the bias term. Note that for $n$ large enough, for $j\in A$ we have
\begin{equation}
\frac{\partial Q(\theta)}{\partial\theta_{j}}  =  -\frac{2}{n}\sum_{i=1}^n(Y_t-\hat{\theta}'_{AL}Z_t)Z_{t}^{(j)}+\frac{\lambda_n}{n}\lambda_{n,j}sign(\hat{\theta}_{AL,j})=0\label{foc}.
\end{equation}
Furthermore, for $n$ large enough $\hat{\theta}_{AL,j}=0$ for $j\notin A$. Thus, we can rewrite the $q$ equations (\ref{foc}) in matrix form
\begin{equation}
0=\frac{2}{n}\sum_{i=1}^n (Y_t-\hat{\theta}_{AL}^{'A}Z_t^{A})Z_t^{A}-\Lambda^{A}_{AL},\label{focm}
\end{equation}
where $\Lambda^{A}_{AL}=(\frac{\lambda_n}{n}\lambda_{n,1}sign(\hat{\theta}_{AL,1}^{A}),\dots,\frac{\lambda_n}{n}\lambda_{n,q}sign(\hat{\theta}_{AL,q}^{A}))'$.
Now consider the term $\frac{1}{n}\sum_{i=1}^n(Y_t-\hat{\theta}'_{AL}Z_t)Z_{t}$. A Taylor expansion around $\theta^{\ast}$ yields
\begin{equation}
\frac{1}{n}\sum_{i=1}^n(Y_t-\hat{\theta}'_{AL}Z_t)Z_{t}=\frac{1}{n}\sum_{i=1}^n(Y_t-\theta'^{\ast}Z_t)Z_{t}-\frac{1}{n}\sum_{i=1}^nZ_tZ_t' (\hat{\theta}_{AL}-\theta^{\ast}).\label{taylor}
\end{equation}
Again, since $\hat{\theta}_{AL,j}=\theta^{\ast}_j=0$ for $j\notin A$ and $n$ large enough, from (\ref{taylor}) it turns out that
\begin{equation}
\frac{1}{n}\sum_{i=1}^n(Y_t-\hat{\theta}'^{A}_{AL}Z_t^{A})Z_{t}^{A}=\frac{1}{n}\sum_{i=1}^n(Y_t-\theta'^{\ast A}Z_t^{A})Z_{t}^{A}-\frac{1}{n}\sum_{i=1}^nZ_t^{A}Z_t'^{A} (\hat{\theta}_{AL}^{A}-\theta^{\ast A}).\label{taylorq}
\end{equation}
Therefore, by combining (\ref{focm}) and (\ref{taylorq}) we have
\begin{equation*}
0=\frac{2}{n}\sum_{i=1}^n(Y_t-\theta'^{\ast A}Z_t^{A})Z_{t}^{A}-\frac{2}{n}\sum_{i=1}^nZ_t^{A}Z_t'^{A} (\hat{\theta}_{AL}^{A}-\theta^{\ast A})-\Lambda^{A}_{AL},
\end{equation*}
i.e.,
\begin{equation*}
\sqrt{n} (\hat{\theta}_{AL}^{A}-\theta^{\ast A})=\left(\frac{1}{n}\sum_{i=1}^nZ_t^{A}Z_t'^{A}\right)^{-1}\left(\frac{1}{\sqrt{n}}\sum_{i=1}^n(Y_t-\theta'^{\ast A}Z_t^{A})Z_{t}^{A}-\frac{\sqrt{n}}{2}\Lambda^{A}_{AL}\right).
\end{equation*}
Since $\frac{\lambda_n}{\sqrt{n}}\to 0$, it turns out that for $j\in A$, $\frac{\lambda_n}{\sqrt{n}}\lambda_{n,j}sign(\hat{\theta}_{AL,j})\to 0$. Therefore,
\[
\sqrt{n}\left(\hat{\theta}_{AL}^{A}-\theta^{\ast A}\right)+\hat{b}_{AL}^{A}\to_d N(0,V^{A}),
\]
where the bias term is given by
\begin{equation*}
\hat{b}_{AL}^{A}=\left(\frac{1}{n}\sum_{i=1}^nZ_t^{A}Z_t'^{A}\right)^{-1}\left(\frac{\lambda_n}{2\sqrt{n}}\lambda_{n,1}^{A}sign(\hat{\theta}_{AL,1}^{A}),\dots,\frac{\lambda_n}{2\sqrt{n}}\lambda_{n,q}^{A}sign(\hat{\theta}_{AL,q}^{A})\right)',
\end{equation*}
This concludes the proof.

\vspace{0.3cm}
\noindent {\bf Proof of Theorem \ref{t2}:}
To prove Theorem \ref{t2},
we use the same arguments as in the proof of Theorem 2 in \citeN{KF00}.
More precisely, let
\[
R_n(u)= \sum_{t=1}^n \left[ (\epsilon_t-u'Z_t/\sqrt{n})^2-\epsilon_t^2\right]+\lambda_n\sum_{i=1}^{p} \lambda_{n,i}\left[ \vert \theta_i^{\ast}+u_i/\sqrt{n}\vert -\vert \theta_i^{\ast}\vert\right].
\]
Note that $R_n$ is minimized at $\sqrt{n}(\hat{\theta}_{AL,\lambda_n}-\theta^{\ast})$. Furthermore,
\begin{eqnarray*}
\sum_{t=1}^n \left[ (\epsilon_t-u'Z_t/\sqrt{n})^2-\epsilon_t^2\right] & \to_d & -2u'W+u'Cu,\\
\lambda_n\sum_{i=1}^{p+r} \lambda_{n,i}\left[ \vert \theta_i^{\ast}+u_i/\sqrt{n}\vert -\vert \theta_i^{\ast}\vert\right] & \to & \sum_{i=1}^{p}\lambda_{0,i}\vert u_i\vert.
\end{eqnarray*}
Thus, $R_n(u)\Rightarrow R(u)$ as $n\to\infty$. Since $R_n$ is convex and $R$ has a unique minimum, it follows from \citeN{G94} that
\[
arg\min(R_n)=\sqrt{n}(\hat{\theta}_{AL,\lambda_n}-\theta^{\ast})\to_d  arg\min(R).
\]
This concludes the proof.

\end{document}